# Highly tunable broadband RF photonic fractional Hilbert transformer based on a Kerr soliton crystal microcomb source

Mengxi Tan, Xingyuan Xu, and David J. Moss

*Abstract* — **We demonstrate an RF photonic fractional Hilbert transformer based on an integrated Kerr microcomb source featuring a record low free spectral range of 48.9 GHz, yielding 75 microcomb lines across the C-band. By programming and shaping the comb lines according to calculated tap weights, we demonstrate that the Hilbert transformer can achieve tunable bandwidths ranging from 1.2 to 15.3 GHz, switchable centre frequencies from baseband to 9.5 GHz, and arbitrary fractional orders. We experimentally characterize the RF amplitude and phase response of the tunable bandpass and lowpass Hilbert transformers with 90 and 45-degree phase shift. The experimental results show good agreement with theory, confirming the effectiveness of our approach as a powerful way to implement the standard as well as fractional Hilbert transformers with broad and switchable processing bandwidths and centre frequencies, together with high reconfigurability and greatly reduced size and complexity.**

## I. INTRODUCTION

The Hilbert transform is a fundamental mathematical function for signal processing systems. It has found wide applications in radar systems, signal sideband modulators, measurement systems, speech processing, signal sampling, and many others [1-3]. Fractional Hilbert transforms can meet specific requirements for secure single sideband communications [2] as well as hardware keys [3] since they provide an additional degree of freedom in terms of a variable phase shift. In practical applications such as multiplexing and demultiplexing signals, analyzing individual sub-channel spectral components, etc., Hilbert transformers are typically realized as a truncated or windowed version of the ideal Hilbert transform impulse response [4-6]. Thus, Hilbert transformers that cover a wide range of different bandpass regions are in high demand. While electronic approaches to Hilbert transformers are subject to the well-known electronic bandwidth bottleneck [7], photonic RF techniques have attracted wide interest due to their numerous advantages, including optical bandwidths that can reach 10's of THz, a strong immunity to electromagnetic interference, and low transmission loss.

A significant amount of research has been devoted to developing and extending the capability of photonic RF Hilbert transformers, including the use of free-space optics [3], although many approaches suffer from a tradeoff between performance and system complexity. Photonic approaches for both standard and fractional Hilbert transforms offer broad operational bandwidths, through the use of novel approaches such as fibre Bragg gratings [8-14], macroing / microdisk resonators [15, 16], and integrated reconfigurable microwave processors [17]. However, most of these approaches focus only on generating the Hilbert transform of the complex optical field, and not the actual RF signal. To implement highly reconfigurable RF photonic Hilbert transformers, transversal schemes have been investigated. Fractional Hilbert transformers based on transversal structures have been demonstrated [18, 19] that offer high reconfigurability. However, methods that employ multiple discrete laser sources present limitations in the overall system footprint, processing performance, and the potential for full monolithic integration.

Recently, [20-39] microcombs have attracted significant interest as a fundamentally powerful tool for microwave processing, due to their ability to offer a large number of coherent wavelengths from one device, thereby increasing the capacity of communications systems [40-42], for example. Further, they have enabled the processing of RF spectra for a wide range of advanced signal processing functions [43-68] as well as neural networks [69-71]. Previously, [43] we reported a Hilbert transformer based on a 200 GHz spaced microcomb generating 20 wavelengths in the C-band that achieved record performance in terms of its RF bandwidth (5 octaves). Subsequently, [63] we demonstrated a fractional Hilbert transformer with a performance in RF bandwidth ranging from 5 to 9 octaves, depending on the fractional order. This was based on a soliton crystal microcomb with an FSR comb spacing of 48.9 GHz, with 17 taps selected on a 200GHz grid from the 75 generated wavelengths across the C-band.[72]

Here, we report a highly reconfigurable and versatile fractional Hilbert transformer based on the same soliton crystal Kerr microcomb source with a 48.9 GHz comb spacing. The device has the ability to switch both its bandwidth and center frequency simultaneously and independently. This versatility together with its high performance is enabled by the use of up to 40 of the 75 comb lines generated by the microcomb over C-

M. Tan and D. J. Moss are with the Optical Sciences Centre, Swinburne University of Technology, Hawthorn, VIC 3122, Australia. (Corresponding e-mail: dmoss@swin.edu.au).

X. Xu is with Electro-Photonics Laboratory, Department of Electrical and Computer System Engineering, Monash University, Clayton, 3800 VIC, Australia band. This represents twice the number of wavelengths, or taps, used in our previous work [63]. We experimentally demonstrate a Hilbert transformer with an RF amplitude and phase response at bandwidths ranging from 1.2 to 15.3 GHz, together with centre frequencies varying from baseband to 9.5 GHz. The experimental results agree well with the theory, confirming the feasibility of our approach towards the realization of high-speed reconfigurable Hilbert transformers with reduced footprint, lower complexity, and potentially reduced cost.

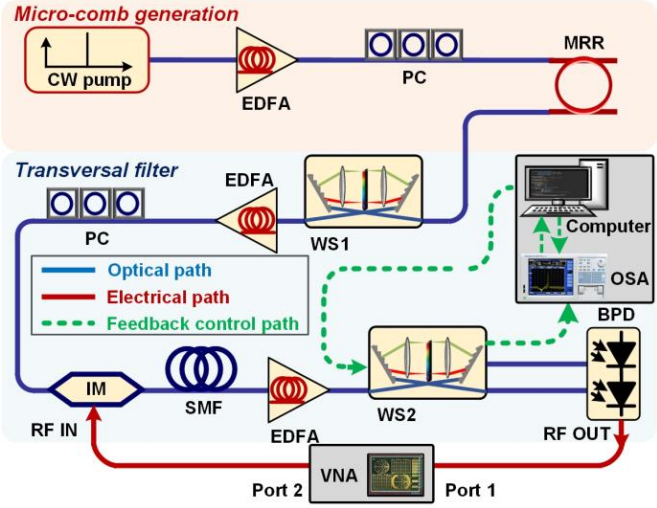

Fig. 1. Schematic diagram of fractional Hilbert transformer based on an integrated Kerr microcomb source. EDFA: erbium-doped fiber amplifier. PC: polarization controller. MRR: micro-ring resonator. WS: WaveShaper. IM: Intensity modulator. SMF: single mode fiber. OSA: optical spectrum analyzer. BPD: Balanced photodetector. VNA: vector network analyzer.

## II. PRINCIPAL OF OPERATION

The spectral transfer function of a general fractional Hilbert transformer is given by [62, 71]:

$$H_P(\omega) = \begin{cases} e^{-j\varphi}, \ if \ 0 \le \omega < \pi \\ e^{j\varphi}, \ if -\pi \le \omega < 0 \end{cases} \quad (1)$$

where $j = \sqrt{-1}$, $\varphi = P \times \pi / 2$ denotes the phase shift, $P$ is the fractional order (when $P = 1$, it becomes a standard Hilbert transformer). The corresponding impulse response is given by a continuous hyperbolic function:

$$h_P(t) = \begin{cases} \frac{1}{\pi t}, t \neq 0 \\ \cot(\varphi), t = 0 \end{cases} \quad (2)$$

This hyperbolic function is truncated and sampled in time with discrete taps for digital implementation. The null frequency is given by:

$$f_c = 1/\Delta t \quad (3)$$

where $\Delta t$ denotes the sample spacing. The coefficient of the tap at $t = 0$ can be adjusted to achieve a tunable fractional order [3]. The normalized power of each comb line is:

$$P_n = \frac{1}{\pi |n - \frac{N}{2} + 0.5|} \quad (4)$$

where $N$ is the number of comb lines, or taps, and $n = 0, 1, 2, \ldots$, $N$-1 is the comb index.

In order to scale the bandwidth of the standard and fractional Hilbert transformer, we multiplied the corresponding impulse response of the spectral transfer function of the Hilbert transformer with a cosine function to vary the operation bandwidth (designed by the Remez algorithm [73]). Hence, the resulting discrete impulse response becomes:

$$h_{TBWHT}(n) = P_n \cdot \cos(2\pi n \cdot f_{BW}) \quad (5)$$

where $f_{BW}$ is the scalable bandwidth. To further switch the centre frequency of the Hilbert transformer, the tap coefficients were multiplied by a sine function to shift the RF transmission spectrum. The corresponding discrete impulse response is given by

$$h_{TCFHT}(n) = P_n \cdot \sin(2\pi n \cdot f_{BW}) \quad (6)$$

Here, we use this transversal approach to achieve Hilbert transformers with both variable bandwidths as well as RF centre frequencies. The transfer function of the transversal structure [75-85] can be described as:

$$F(\omega) = \sum_{n-0}^{M-1} h(n)e^{-j\omega nT} \quad (7)$$

where $M$ is the number of taps, $\omega$ is the RF angular frequency, $T$ is the time delay between adjacent taps, and $h(n)$ is the tap coefficient of the $n^{th}$ tap.

Figure 1 illustrates the schematic diagram of the microcomb-based RF photonic Hilbert transformer. The microcomb was generated by pumping a nonlinear high-Q MRR, fabricated on a high-index doped silica glass platform, with a CW laser, amplified with an erbium-doped fibre amplifier (EDFA), with the polarization state aligned to the TE mode of the MRR. When the pump wavelength was swept manually across one of the MRR's resonances and the pump power was high enough to generate sufficient parametric gain, optical parametric oscillation occurred, ultimately generating Kerr microcomb with a spacing equal to the free spectral range of the MRR. The generated microcomb was then spectrally flattened and modulated by the WaveShapers to achieve the designed tap weights.

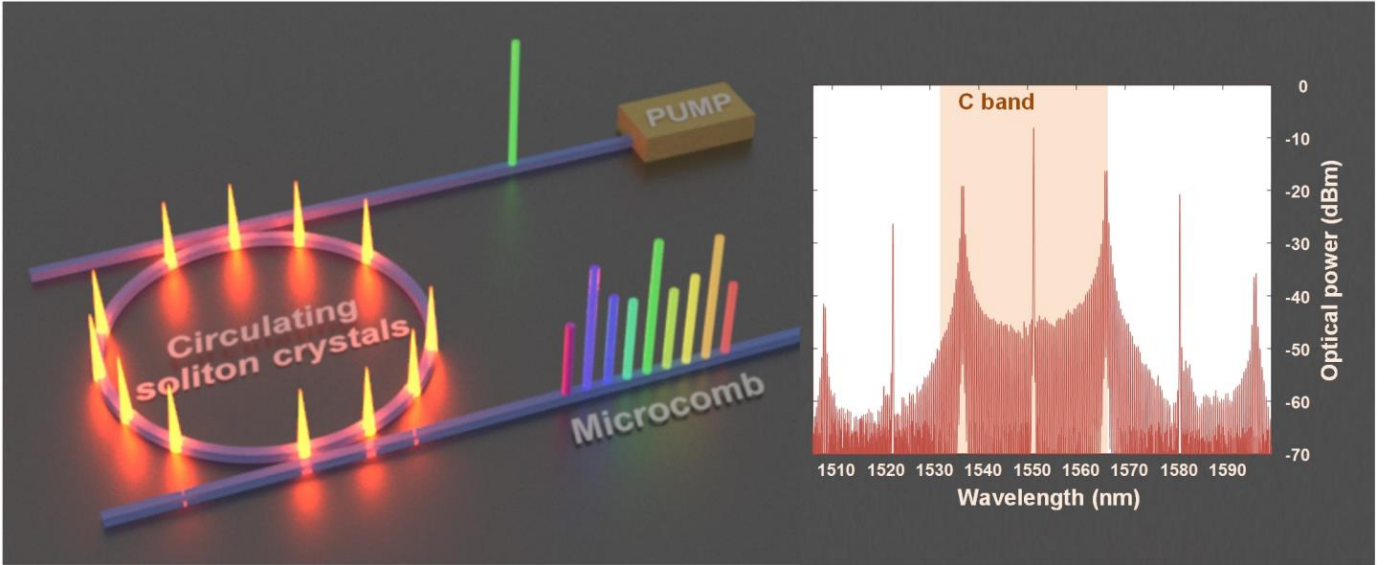

Fig. 2. Schematic illustration of the integrated MRR for generating the Kerr frequency comb and the optical spectrum of the generated soliton crystal combs with a 100-nm span.

Following this, the shaped comb lines were then imprinted with the RF signal via an intensity modulator, yielding replicas of the RF input waveform in the optical domain, transmitting the wavelength through a dispersive medium to acquire wavelength-dependent delays. The weighted and delayed signals were summed together upon photodetection and converted back into the RF domain.

## III. EXPERIMENTAL RESULTS

The MRRs used here were fabricated on a platform based on Hydex glass [30, 33, 34, 38, 39] with CMOS compatible fabrication processes. First, Hydex films ($n = \sim1.7$ at 1550 nm) were deposited using PECVD, then patterned by deep ultraviolet (UV) photolithography and etched via reactive ion etching [29] to achieve waveguides with very low surface roughness. Finally, an upper cladding layer composed of silica

($n = \sim1.44$ at 1550 nm) was deposited. We often employ a vertical coupling scheme between the ring resonator and bus, where the gap (approximately 200 nm) can be controlled by film growth, much more accurate than using lithographic techniques. The advantages of our platform, particularly in relation to optical microcombs, include ultra-low linear loss ($\sim0.06$ dB · cm$^{-1}$), a moderately high optical nonlinear parameter ($\sim233$ W$^{-1}$ · km$^{-1}$), and in particular, a negligible nonlinear loss - even up to extremely high intensities of about 25 GW · cm$^2$. As a result of the ultra-low loss of the platform, our MRR features narrow resonance linewidths corresponding to Q factors of up to $\sim1.5$ million for the 48.9 GHz MRR. The insertion loss of the through-port was as low as 0.5 to 1 dB/facet – a result of very efficient on-chip mode converters that allowed packaging the device with fibre pigtails. The radius of the MRR was $\sim592$ μm, corresponding to an optical FSR of 0.393 nm or 48.9 GHz. This small FSR greatly increased the number of wavelengths (channels) available over the C band to as many as

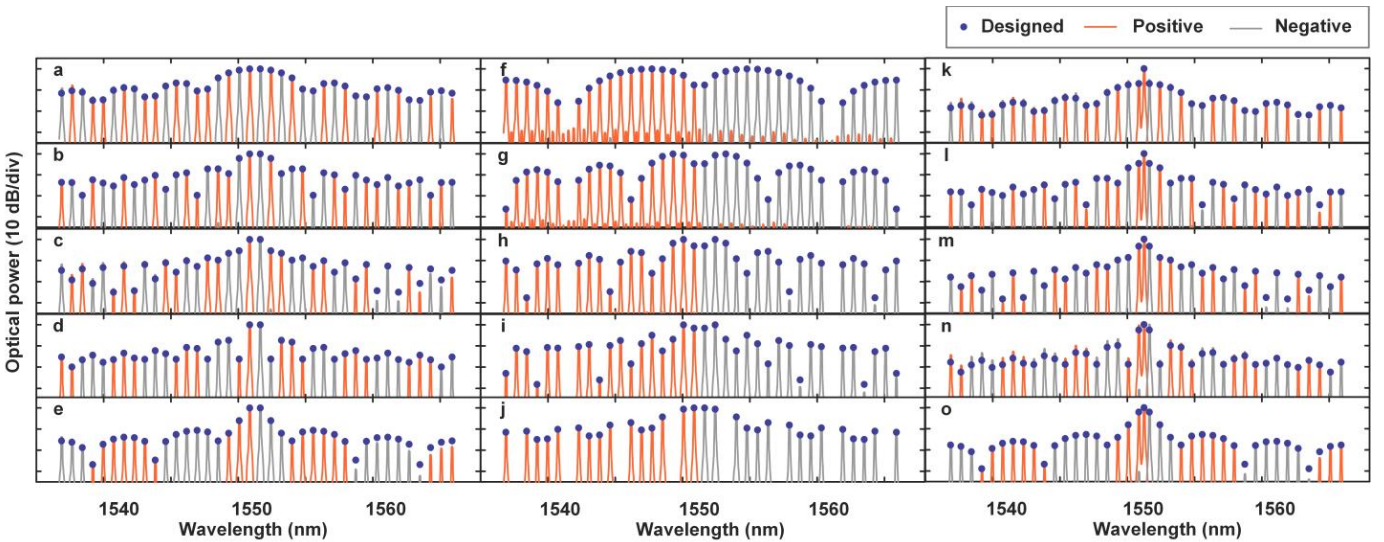

Fig. 3. Designed and measured optical spectra for (a)~(e) Tunable bandpass Hilbert transformer with 90-degree phase shift. (f)~(j) Tunable lowpass Hilbert transformer with 90-degree phase shift. (k)~(o) Tunable bandpass fractional Hilbert transformer with 45-degree phase shift. Note that all taps are on a ~ 100GHz grid (or 2 x 48.9 = 97.8 GHz) except for the central 3 taps of the fractional transformer (right, (k)~(o)) that are spaced by 48.9GHz.

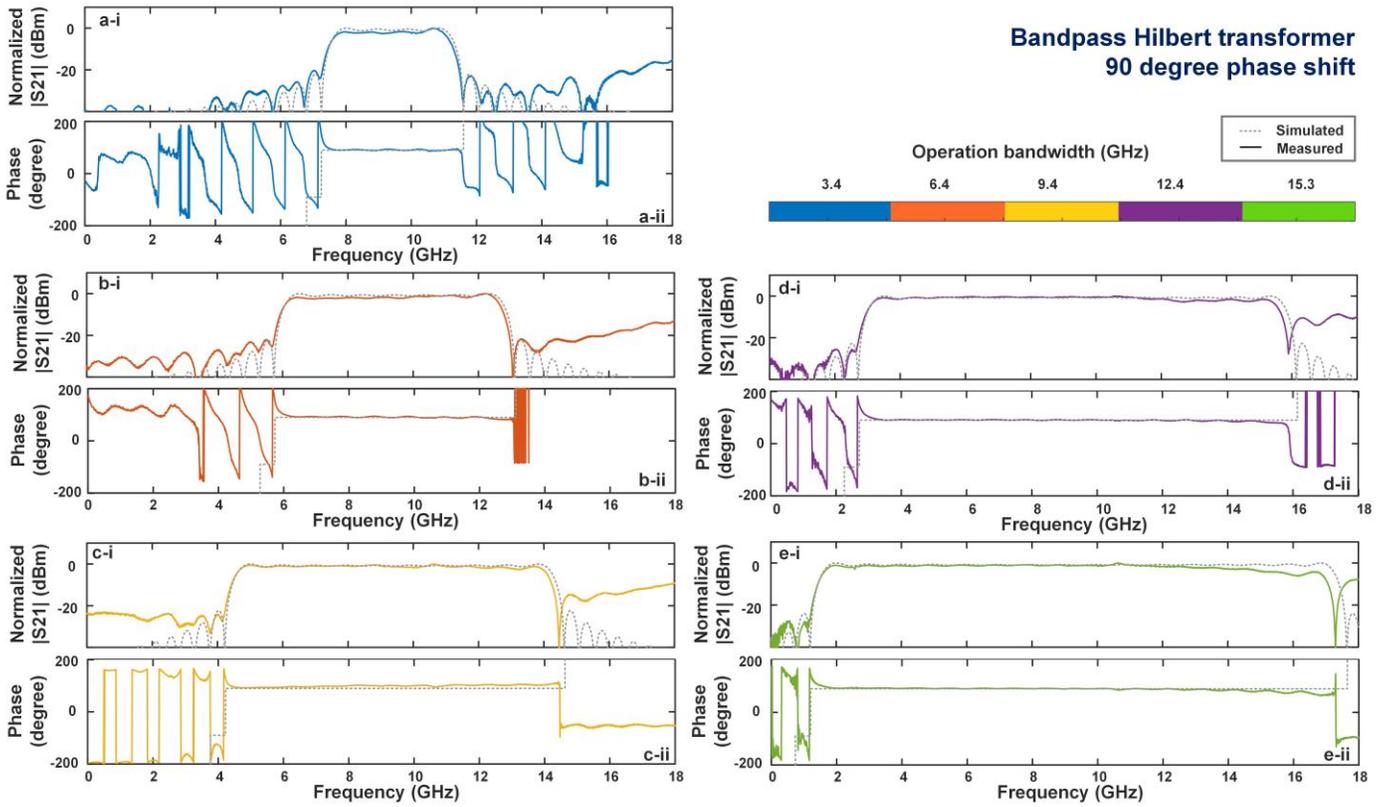

Fig. 4. Simulated (dashed curves) and experimental (solid curves) results of tunable bandpass Hilbert transformer with 90-degree phase shift. Note that the resulting bandwidths are colour coded corresponding to the graph colours (top right legend) and correspond to a) blue 3.4 GHz, b) red 6.4 GHz, c) yellow 9.4 GHz, d) purple 12.4 GHz, and e) green 15.3 GHz.

75 wavelengths to be used for the RF device. This is more than a factor of two higher than our previous work [63].

To generate microcombs with the 48.9 GHz device, the CW pump power was amplified to ~30.5 dBm, and the wavelength swept from blue to red near one of the TE resonances of the MRR at ~1553.2 nm [51]. When the detuning between the pump wavelength and cold resonance of the MRR became small enough, such that the intra-cavity power reached a threshold, modulation instability (MI) driven oscillation occurred [22]. Primary combs were then generated with the spacing determined by the MI gain peak – mainly a function of the dispersion and intra-cavity power. As the detuning was changed further, distinctive 'fingerprint' optical spectra eventually appeared (Fig. 2). The spectra are similar to what has been reported from spectral interference between tightly packed solitons in the cavity – so-called "soliton crystals" [26-28]. The soliton crystal states provided the lowest RF noise states of all our microcombs, and have been used as the basis for an RF oscillator with low phase-noise [57].

The soliton crystal microcomb was first flattened via an optical spectral shaper (Finisar, WaveShaper 4000S) and then modulated with the RF input signal to multicast the RF waveform onto all wavelength channels simultaneously, yielding 75 replicas, although only 38 or 39 taps were actually used, on a 98 GHz spacing (2 x 48.9 GHz) for the device. These were then passed through ~3.84 km of standard single mode fibre (SMF) to provide the progressive time delays between wavelengths. The fibre was approximately twice as long as that used in [63] in order to yield comparable RF bandwidths. The dispersion of the SMF was ~17.4 ps/nm/km, corresponding to a time delay $\Delta t = 26.25$ ps between adjacent wavelengths. Next, the second WaveShaper accurately shaped the comb power according to the designed tap coefficients, with the shaped comb spectrum shown in Fig. 3, for the integral order (90 degree phaseshift) tunable bandpass (a)~(e), integral order (90 degree phaseshift) tunable lowpass (f)~(j), and tunable fractional order (45 degree phaseshift) bandpass filter (k)~(o). Note that all devices used a ~100GHz tap spacing (or more precisely 48.9 x 2 = 97.8 GHz), yielding 39 tap lines across the C-band. Note the fractional order device used an extra single tap line at the centre wavelength, yielding 40 wavelengths overall, with the centre 3 wavelengths then being spaced at 48.9 GHz. The result was that the Nyquist zone for the fractional device was 24.5 GHz rather than 48.9 GHz for the integral order devices. For standard Hilbert transformer, this extra specific tap coefficient is not needed and so in principle the full 48.9 GHz comb (75 lines over the C-band) could have been used although these results are not shown here.

The wavelength channels for both positive (solid red line) and negative (solid gray line) taps were separately measured by an optical spectrum analyser (OSA), achieving good performance that agreed well with the theory (blue dot). Finally, the weighted and delayed replicas were combined and converted back into the RF domain via a balanced photodetector (Finisar BPDV2150R).

The system RF frequency response was characterized with a calibrated vector network analyser (VNA, Agilent MS4644B) to measure the RF transmission and phase response. Fig. 4 (a-

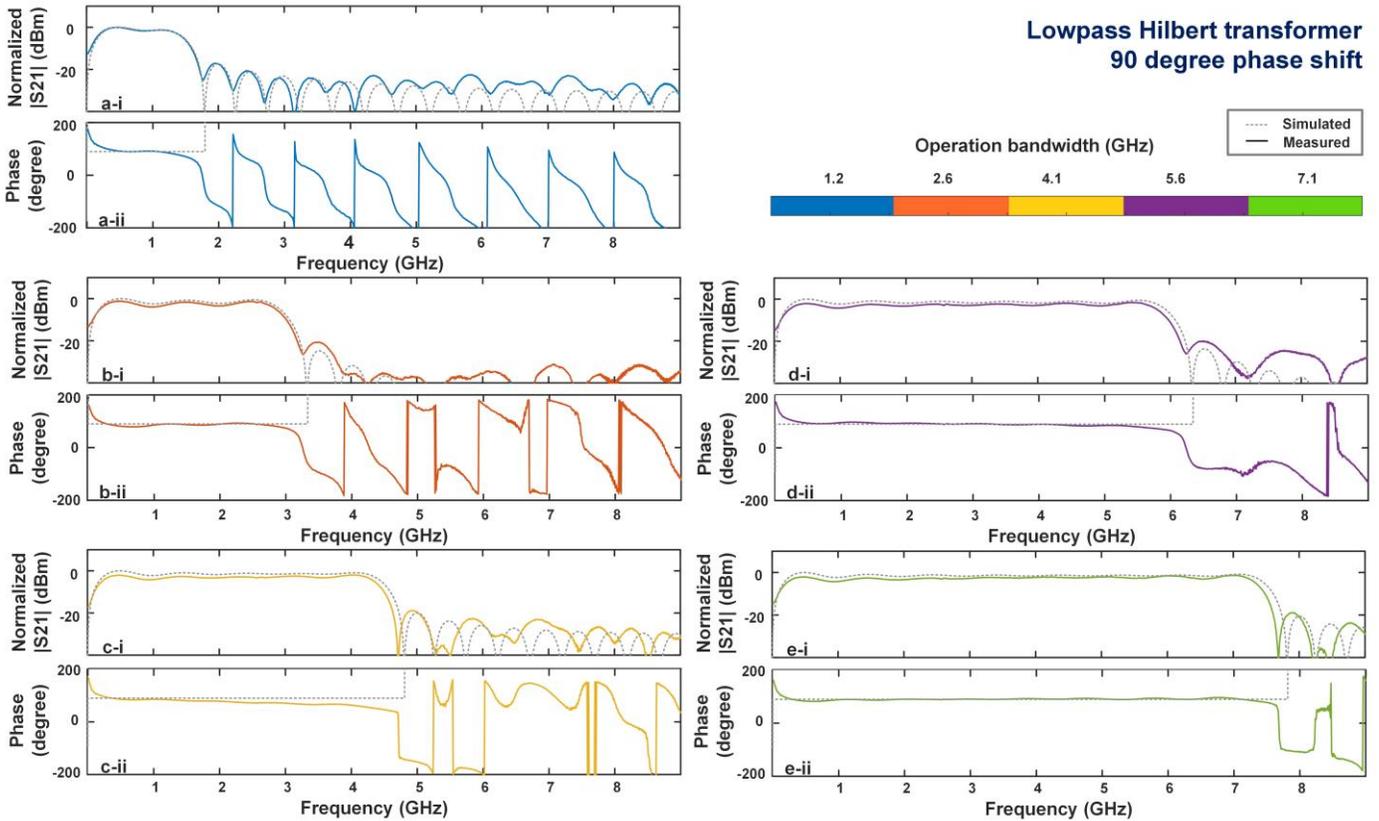

Fig. 5. Simulated (dashed curves) and experimental (solid curves) results of tunable lowpass Hilbert transformer with 90-degree phase shift. Note that the resulting bandwidths are colour coded corresponding to the graph colours (top right legend) and correspond to a) blue 1.2 GHz, b) red 2.6 GHz, c) yellow 4.1 GHz, d) purple 5.6 GHz, and e) green 7.1 GHz.

e) presents the simulated (dashed curves) and measured (solid curves) RF frequency response for both the magnitude and phase of the standard Hilbert transformer, yielding variable bandwidths ranging from 3.4 to 15.3 GHz. The centre frequency of the Hilbert transformer was set into half of the $FSR_{RF}$, which was $19/2 = 9.5$ GHz in our case.

Figure 5 shows the measured results for the RF amplitude and phase response of the lowpass Hilbert transformer, showing tunable bandwidths ranging from 1.2 to 7.1 GHz that match closely with the simulated results. We also performed a demonstration of a fractional Hilbert transformer with switchable RF bandwidths ranging from 3.5 to 15.2 GHz. The simulated and measured RF amplitude and phase responses are shown in Fig. 6. We achieved a fractional order of 0.5, which corresponds to a 45-degree phase shift, using up to 39 wavelengths or taps. This increase of a factor of two from our previous work resulted in a significant reduction in the overall root-mean-square error (RMSE) as well as a modest increase in the potential operational bandwidth, as was predicted [63]. The increased number of taps used here compared with our previous work (39 here versus 17 taps in [63]) only resulted in an increase in overall bandwidth of about 0.6 GHz, although the performance in octaves was improved a bit more than this (5 octaves for 17 taps versus 6.3 octaves achieved here, although the results from [63] are experimental – see Table I).

## IV. CONCLUSION

We demonstrate a broadband RF photonic Hilbert transformer with variable bandwidths as well as RF centre frequency, based on a Kerr soliton crystal microcomb. Up to 39 wavelengths, or taps were used, resulting in a tunable bandwidth ranging from 1.2 to 15.3 GHz as well as a switchable centre frequency from baseband to 9.5 GHz. Dynamic tuning of the bandwidth, as well as the centre frequency, was achieved by programming and adjusting the tap weights. This microcomb based approach represents a solid step towards achieving a fully integrated photonic signal processor for future ultra-high-speed RF systems.

Table. I
Calculated performance of the standard Hilbert transformer

|  | 4 taps | 16 taps | 38 taps | 80 taps |
|---|---|---|---|---|
| Lower cutoff frequency (GHz) | 16.3 | 17.9 | 18.2 | 18.4 |
| Upper cutoff frequency (GHz) | 2.2 | 0.5 | 0.2 | 0.1 |
| 3-dB bandwidth (GHz) | 14.1 | 17.4 | 18.0 | 18.2 |
| Octaves | 2.9 | 5.0 | 6.3 | 7.4 |

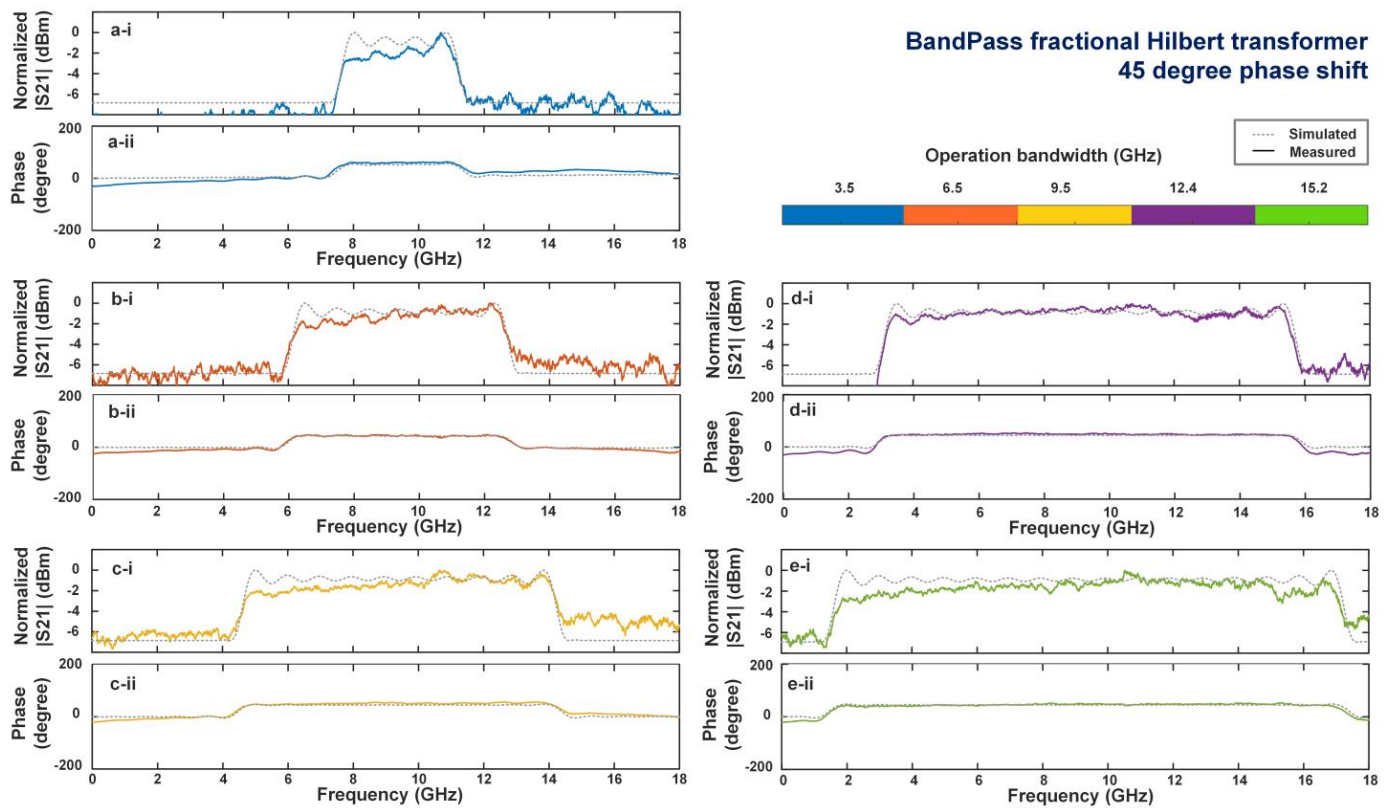

Fig. 6. Simulated (dashed curves) and experimental (solid curves) results of tunable bandpass fractional Hilbert transformer with 45-degree phase shift. Note that the resulting bandwidths are colour coded corresponding to the graph colours (top right legend) and correspond to a) blue 3.5 GHz, b) red 6.5 GHz, c) yellow 9.5 GHz, d) purple 12.4 GHz, and e) green 15.2 GHz.